\documentclass[aps,pra,floatfix,amsmath,amssymb,twocolumn]{revtex4}

\usepackage[american]{babel}
\usepackage{graphicx}
\usepackage{dcolumn}
\usepackage{bm}
\usepackage{hyphenat}
\usepackage{amssymb}   
\usepackage{verbatim}
\usepackage{xpatch}
\usepackage{color}
\usepackage{epstopdf}

\usepackage{hyperref}

\begin{document}

\title{Interrogating the temporal coherence of EUV frequency combs with\\highly charged ions}

\author{Chunhai Lyu}
\altaffiliation{chunhai.lyu@mpi-hd.mpg.de}
\affiliation{Max-Planck-Institut f\"{u}r Kernphysik, Saupfercheckweg 1, 69117 Heidelberg, Germany}
\author{Stefano M. Cavaletto}
\altaffiliation[Current address: ]{Department of Chemistry and Department of Physics and Astronomy, University of California, Irvine, CA 92697-2025, USA}
\affiliation{Max-Planck-Institut f\"{u}r Kernphysik, Saupfercheckweg 1, 69117 Heidelberg, Germany}
\author{Christoph H. Keitel}
\affiliation{Max-Planck-Institut f\"{u}r Kernphysik, Saupfercheckweg 1, 69117 Heidelberg, Germany}
\author{Zolt\'{a}n Harman}
\affiliation{Max-Planck-Institut f\"{u}r Kernphysik, Saupfercheckweg 1, 69117 Heidelberg, Germany}

\date{\today}

\begin{abstract}

A scheme to infer the temporal coherence of EUV frequency combs generated from intra-cavity high-order harmonic generation is put forward. The excitation dynamics of highly charged Mg-like ions, interacting with EUV pulse trains featuring different carrier-envelope-phase fluctuations, are simulated. While demonstrating the microscopic origin of the macroscopic equivalence between excitations induced by pulse trains and continuous-wave lasers, we show that the
coherence time of the pulse train can be determined from the spectrum of the excitations. The scheme will provide a verification of the comb temporal coherence at time scales several orders of magnitude longer than current state of the art, and at the same time will enable high-precision spectroscopy of EUV transitions with a relative accuracy up to $\delta\omega/\omega\sim10^{-17}$.

\end{abstract}

\maketitle

%

A train of evenly delayed coherent electromagnetic pulses resembles a structure in the frequency domain with uniformly displaced frequency peaks, i.e, a frequency comb~(FC)~\cite{cundiff2003colloquium,ye2005femtosecond}. The inverse of the coherence time $\tau_{\rm{c}}$ of such a FC determines the width of each comb tooth, and can be inferred by measuring the beating notes between the corresponding pulse train and an independent ultrastable continuous-wave (cw) reference laser~\cite{schibli2008optical,benko2012full}. For optical FCs, coherence times longer than 1~s, or tooth widths narrower than 1~Hz, have been measured~\cite{schibli2008optical,benko2012full}, which allows wide applications of FCs in high-precision spectroscopy~\cite{picque2019frequency,fortier201920}, the search for exoplanets~\cite{li2008laser,steinmetz2008laser} and the construction of ultrastable optical atomic clocks~\cite{ludlow2015optical}.

Through intra-cavity high-order harmonic generation (HHG)~\cite{mills2012xuv} of femtosecond infrared (IR) pulse trains, coherent extreme-ultraviolet (EUV) pulse trains representing EUV FCs have been demonstrated~\cite{gohle2005frequency,jones2005phase}. This could allow high-precision spectroscopy in the EUV regime~\cite{cingoz2012direct,dreissen2019high} and enable next-generation atomic clocks based on EUV transitions~\cite{adams2013x,cavaletto2014broadband,seiferle2019energy}. However, due to the lack of cw EUV reference lasers, the temporal coherence of an EUV FC is mainly investigated by splitting the EUV pulse train into two pathways and then recombing them to perform Michelson interference~\cite{jones2005phase,yost2009vacuum}. The observed cross correlation between two adjacent pulses reveals the well-defined temporal coherence on the time scale of 10~ns~\cite{yost2009vacuum}. This result was further verified by the direct frequency-comb spectroscopy (DFCS) of atomic transitions in Ne and Ar~\cite{cingoz2012direct}: the measured fluorescence spectra exhibited a full-width at half-maximum (FWHM) of 10~MHz, implying that the coherence time of the EUV FC is longer than 16~ns. 

Instead of splitting the EUV pulse train, the IR pulse train can be split and sent into two isolated cavities where HHG takes place separately~\cite{benko2014extreme}. The cross-correlation measurement of the two almost independently generated EUV pulse trains indicates that HHG itself may be the leading process affecting the coherence time of the EUV frequency combs~\cite{benko2014extreme}. This suggests that when the IR frequency comb is locked to a mHz ultrastable cw laser~\cite{schibli2008optical,matei20171.5um}, EUV FCs with coherence times of $\tau_{\rm{c}}\gtrsim1$~s (tooth width $\lesssim160$~mHz) could be obtained. However, recent studies~\cite{corsi2017ultimate,eramo2018analytic} argue that such a fine comb structure may not be achieved with currently available feedback loops, and have set an ultimate upper limit on the comb coherence time of EUV FCs of $\tau_{\rm{c}}\lesssim64$~ns (tooth width $\gtrsim2.5$~MHz). 
Therefore, verifying the coherence time of the EUV frequency comb at longer time scales becomes essential. Currently, this is limited either by the longest arm length tunable in the Michelson-interference schemes~\cite{jones2005phase,yost2009vacuum} or by the longest lifetimes of the EUV transitions available in the DFCS schemes~\cite{cingoz2012direct}. 

Highly charged ions (HCIs) can be produced, e.g., in an electron-beam ion trap (EBIT)~\cite{micke2018heidelberg} and then be moved to a cryogenic Paul trap (CryPT)~\cite{schwarz2012cryogenic,leopold2019cryogenic,micke2020coherent} 
for interactions with external lasers~\cite{nauta2017towards,micke2020coherent,nauta2020100MHz}. Due to the existence of environment-insensitive forbidden optical transitions, HCIs are of great interest in frequency metrology and for tests of fundamental physics~\cite{lopez2016frequency,kozlov2018highly}. By employing excited configurations, specific HCIs also provide forbidden transitions that can be probed by EUV frequency combs. This would enable the detection of the coherence time of EUV pulse trains at time scales longer than 100~ns, and at the same time render high-precision spectroscopy of EUV transitions possible.

\begin{table}[b]
\caption{Transition energy $\hbar\omega$ and lifetime $\tau$ of the [Ne]$3s3p$ states in Mg-like ions. The lifetimes for Ca$^{8+}$ and Ti$^{10+}$ are adopted from ref.~\cite{gustafsson2017mcdhf}}
\begin{tabular}{c|cc|cc|cc|cc}
\hline
\hline
                & \multicolumn{2}{c}{$^1P_1$} & \multicolumn{2}{c}{$^3P_2$} &\multicolumn{2}{c}{$^3P_1$} &\multicolumn{1}{c}{$^3P_0$} \\
ions            & $\hbar\omega$ (eV) & $\tau$  & $\hbar\omega$ (eV) & $\tau$  & $\hbar\omega$ (eV) & $\tau$  & $\hbar\omega$ (eV)  \\
\hline
S$^{4+}$       & 15.765             & 301~ps   & 10.434             & 16~s    & 10.339             & 6.7 $\mu$s        & 10.294                  \\
\hline
Ar$^{6+}$      & 21.167             & 123~ps   & 14.331             & 5.3~s    & 14.122             & 1.3 $\mu$s        & 14.023                  \\
\hline
Ca$^{8+}$      & 26.592             & 94~ps    & 18.339             & 2.2~s    & 17.937             & 0.4 $\mu$s        & 17.752                  \\
\hline
Ti$^{10+}$     & 32.108             & 72~ps    & 22.487             & 0.4~s    & 21.790             & 0.1 $\mu$s        & 21.476                 \\
\hline
\hline
\end{tabular}
\label{Mg-like-ions}
\end{table}

In this Letter, we put forward the interrogation of the coherence time of an EUV FC with highly charged Mg-like ions featuring a ground-state configuration of [Ne]$3s^{2}~^1S_0$. The energies and lifetimes of the [Ne]$3s3p$ excited-state configurations for selected ions are presented in Table~\ref{Mg-like-ions}. These values are calculated employing multiconfiguration Dirac--Hartree--Fock (MCDHF) theory~\cite{fischer2019grasp2018,gustafsson2017mcdhf} and referenced with the experimental values available from the NIST atomic database~\cite{kramida2019nist}. In contrast to the EUV transitions in neutral atoms that usually decay within 100~ns~\cite{cavalieri2002ramsey,cingoz2012direct,kandula2010extreme,witte2005deep,altmann2018deep,dreissen2019high}, 
the EUV transitions in Table~\ref{Mg-like-ions} possess lifetimes around both 1~$\mu$s and 1~s. Therefore, they can be used to investigate the coherence time of an EUV pulse train for harmonics from the 9th to the 19th order and beyond. By extending the light--matter interaction to account for phase fluctuations in the pulse train, we show that the coherence time can be determined either through DFCS, where millions of pulses interact with the ion~\cite{cingoz2012direct,felinto2009theory}, or via Ramsey frequency comb spectroscopy (RFCS)~\cite{cavalieri2002ramsey,witte2005deep,morgenweg2014ramsey,altmann2018deep,dreissen2019high}, 
where only two pulses separated in time interact with the ion.

\begin{figure}[t]
\includegraphics[width=0.45\textwidth]{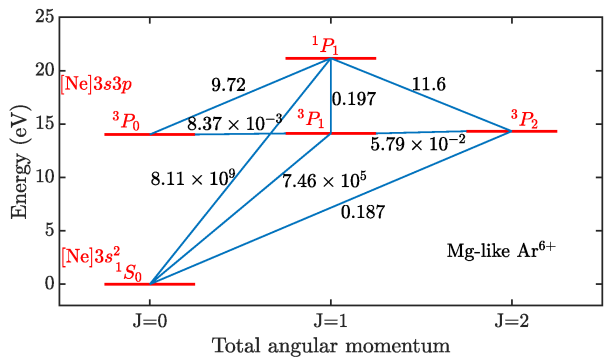}\\
\caption{\label{level-Ar6} Level structure and radiative transition rates (in units of per second) of Ar$^{6+}$.}
\end{figure}

\textit{Mg-like Ar$^{6+}$ --} 
We consider Mg-like Ar$^{6+}$ ions as shown in Fig.~\ref{level-Ar6}: the $^1P_1$ state decays to the ground state through a fast $E1$ transition within one cycle of the EUV pulse, while the $^3P_1$ and $^3P_2$ states can effectively interact with hundreds and millions of pulses before they decay, and thus interrogate the temporal coherence of the EUV pulse trains at time scales around 1.3~$\mu$s and 5.6~s, respectively. State-of-the-art experimental energies~\cite{trigueiros1997atomic} of the transitions from the $^3P_1$ and $^3P_2$ states to the ground state are 14.12248(24)~eV and 14.33133(25)~eV, respectively, with a relative uncertainty of $\delta\omega/\omega\sim1.7\times10^{-5}$. We will show that the investigations of the FC coherence time can lead to a reduction of this uncertainty by several orders of magnitude.

\begin{figure*}[t]
\includegraphics[width=0.95\textwidth]{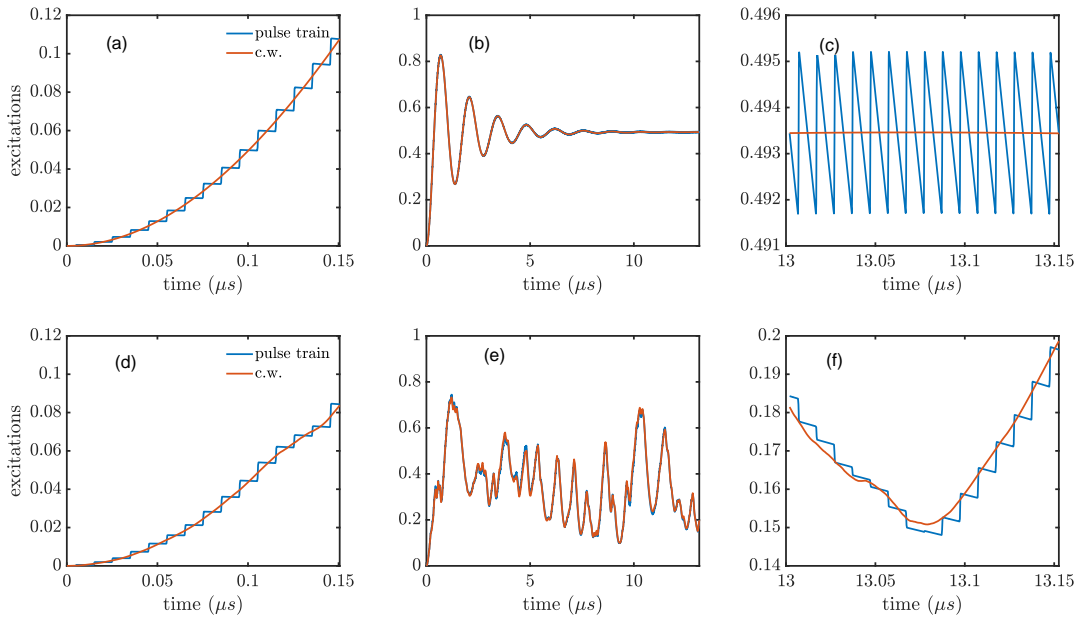}\\
\caption{\label{average_power_4mW}The population $\rho_{\rm{ee}}(t)$ of the $^3P_1$ state in Ar$^{6+}$ induced by two $4$-mW EUV FCs. (a-c) comb~1 with $\tau_{\rm{c}}=1$~s~\cite{benko2014extreme}. (d-f) comb~2 with $\tau_{\rm{c}}=64$~ns~\cite{corsi2017ultimate}. 
(a,d) and (c,f) show the excitations during the first and last 15 pulses in (b,e), respectively.}
\end{figure*}

\textit{Frequency comb --} 
In the time domain, a FC is described as a train of consecutive pulses with a repetition time of $T_{\rm{r}}$~\cite{cundiff2003colloquium,ye2005femtosecond}: 
\begin{eqnarray} 
E(t)&=&E_{\text{p}}~\sum_{j}f(t-jT_{\rm{r}})\text{cos}(\omega_0t+\phi(t)),
\end{eqnarray}
where $E_{\text{p}}$ is the peak strength of the electric field, $f(t-jT_{\rm{r}})$ is the normalized envelope of the $j$-th pulse under a carrier frequency of $\omega_0$, and $\phi(t)$ is the carrier-envelope phase (CEP) at time $t$. For an ideal case where all these parameters are stable and deterministic, one obtains an infinitely correlated pulse train with a perfect comb structure in the frequency domain. However, fluctuations in $T_{\rm{r}}$ and $\phi(t)$ lead to a finite correlation time that broadens the lineshape of each tooth~\cite{corsi2017ultimate,eramo2018analytic,endo2018ultralow,bartels2004stabilization,liehl2019deterministic}. 
Here, we only consider the CEP fluctuations which we model as a random walk process such that~\cite{corsi2017ultimate,eramo2018analytic} 
\begin{eqnarray}
\phi(t)&=&\int_{0}^ts(t')dt', 
\end{eqnarray}
where $s(t)$ represents a Gaussian white noise with autocorrelation $\left<s(t)s(t')\right>=\sigma^2\delta(t-t')$. 
This results in a coherence time of $\tau_{\rm{c}}=1/2\pi\sigma^2$, corresponding to a tooth FWHM of $\sigma^2$~\cite{eramo2018analytic}.

\textit{Bloch equations --} 
We provide quantum dynamical simulations of the excitations of Ar$^{6+}$ ions coupled to an EUV pulse train. 
The duration of each pulse is assumed to be 200~fs with a repetition time of $T_{\rm{r}}=10$~ns~\cite{nauta2017towards}, corresponding to a FWHM bandwidth of 2.19~THz and repetition rate of 100~MHz. 
The carrier frequency $\omega_0$ is tuned to the $^3P_1\rightarrow{}^1S_0$ transition. 
The energy separations between the levels shown in Table~\ref{Mg-like-ions} are much larger than the bandwidth of the frequency comb. Therefore, the ions can be modeled as two-level systems whose dynamics are described by Bloch equations~\cite{temkin1993excitation,ziolkowski1995ultrafast,vitanov1995coherent,scully1999quantum} in the rotating-wave approximation: 
\begin{eqnarray}
\dot{\rho}_{\text{ee}}&=&-\text{Im}\left[\mu^*\mathcal{E}^*(t)\rho_{\text{eg}}(t)\right]-\Gamma\rho_{\text{ee}}(t),
\label{drhoeeapp-ss}\\
\dot{\rho}_{\text{gg}}&=&\text{Im}\left[\mu^*\mathcal{E}^*(t)\rho_{\text{eg}}(t)\right]+\Gamma\rho_{\text{ee}}(t),
\label{drhoggapp-ss}\\
\dot{\rho}_{\text{eg}}&=&\frac{i\mu\mathcal{E}(t)}{2}\left[\rho_{\text{ee}}(t)-\rho_{\text{gg}}(t)\right]+\left(i\Delta-\frac{\Gamma}{2}\right)\rho_{\text{eg}}(t).
\label{drhoegapp-ss}
\end{eqnarray} 
Here, $\rho_{\text{ee}}(t)$ and $\rho_{\text{gg}}(t)$ are the populations of the excited and ground states, respectively. $\rho_{\text{eg}}(t)=\rho^*_{\text{ge}}(t)$ is the off-diagonal element of the density matrix. $\mu$ is the dipole moment that couples to the field envelope 
\begin{eqnarray}
\mathcal{E}(t)&=&E_{\text{p}}~\sum_{j}f(t-jT_{\rm{r}})e^{i\phi(t)}.
\end{eqnarray} 
Furthermore, $\Gamma$ is the spontaneous-emission rate and $\Delta=\omega_0-\omega$ is the detuning. 

\textit{Population dynamics --} 
Though most EUV FCs have an average power around tens of~$\mu$W~\cite{mills2012xuv}, an average power of several mW has been achieved recently~\cite{porat2018phase}. Figure~\ref{average_power_4mW} shows the excitation dynamics by two 4-mW combs with different coherence times: while Figs.~\ref{average_power_4mW}a-c refer to a comb with $\tau_{\rm{c}}=1$~s as in ref.~\cite{benko2014extreme}, \mbox{Figs.~\ref{average_power_4mW}d-f} stand for the comb from ref.~\cite{corsi2017ultimate} with $\tau_{\rm{c}}=64$~ns. The EUV light is supposed to be focused onto a 10-$\mu\text{m}^2$ spot such that $E_{\text{p}}=1.19\times10^{8}~\text{V/m}$. The excitations (red lines) induced by a 170-nW resonant cw laser, with the same fluctuating phase $\phi(t)$ but a constant field strength of $\mathcal{E}_{\rm{eff}}=\int_{0}^{T_{\rm{r}}}\mathcal{E}(t)dt/T_{\rm{r}}=3580~\text{V/m}$, are also shown. This cw laser, featuring a Rabi frequency of 720~kHz, bears the same power as the average power held by a single comb mode at $\omega_0$. 

For both combs, one obtains coherent accumulations~\cite{felinto2001accumulative,felinto2003coherent,marian2004united,aumiler2009time} 
of stepwise excitations (Figs.~\ref{average_power_4mW}a,d). The amount of each stepwise excitation within the 200-fs pulse duration is equivalent to the amount of continuous excitation by the corresponding cw laser within a period of $T_{\rm{r}}=10$~ns, representing the microscopic origin of the macroscopic equivalence~\cite{felinto2009theory} between the pulse-train and cw-laser excitations illustrated in Figs.~\ref{average_power_4mW}b,e. While the similarities in the excitations by the first 15 pulses shown in Figs.~\ref{average_power_4mW}a,d are clearly apparent for the two combs, differences start to emerge at times beyond the 64-ns-long coherence time of comb~2. For comb~1, whose CEP dephasing is negligible, the excitation by each pulse adds up coherently and induces the Rabi oscillation~\cite{allen1987optical} shown in Figs.~\ref{average_power_4mW}a,b. For comb~2, however, the CEP dephasing starts to slow down the excitations, and a chaotic evolution is observed at long time scales (Fig.~\ref{average_power_4mW}d,e). 

Furthermore, when the time becomes much longer than the 1.3-$\mu$s excited-state lifetime, the coherent excitation of comb~1 evolves into a dynamical steady state~\cite{moreno2014comparative,yudin2016dynamic,cavaletto2014Quantum}. 
The population decayed during the absence of the pulse within each cycle (blue line in Fig.~\ref{average_power_4mW}c) is subsequently re-pumped by the next pulse, revealing the distinct transient behavior in comparison to the constant population (red line) induced by cw lasers. The excitation dynamics of comb~2, however, are always random and do not approach any steady state (Fig.~\ref{average_power_4mW}f).

\begin{figure}[t]
\includegraphics[width=0.45\textwidth]{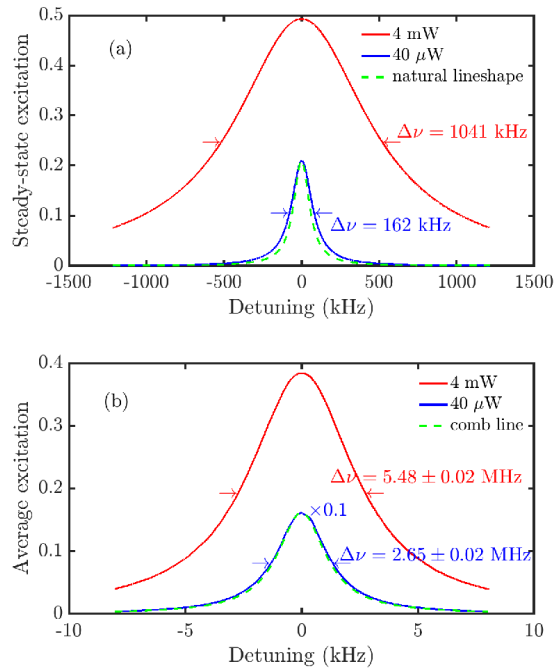}\\
\caption{\label{excitation_vs_detuning} Excitations vs detuning. (a) steady-state excitations for comb~1 with $\tau_{\rm{c}}=1$~s~\cite{benko2014extreme}; (b) 1.3-ms-averaged excitations for comb~2 with $\tau_{\rm{c}}=64$~ns~\cite{corsi2017ultimate}. For both cases, red and blue lines refer to 4-mW and 40-$\mu$W comb powers, respectively. The green dashed lines represent (a) the Lorentzian lineshape of the 122-kHz-wide $^3P_1\rightarrow{}^1S_0$ ionic transition and (b) the 2.50-MHz-wide comb tooth.}
\end{figure}

\textit{DFCS scheme --} 
To determine the coherence time of the FCs and the energy of the ionic transition, Fig.~\ref{excitation_vs_detuning} illustrates the excitations as a function of the detuning $\Delta$. While Fig.~\ref{excitation_vs_detuning}a represents the steady-state excitation spectra for comb~1, the spectra in Fig.~\ref{excitation_vs_detuning}b for comb~2 
are the average excitations over a duration of 1.3~ms. The results for FCs with an average power of 40-$\mu$W (blue lines) are also presented. For comb~1, whose 160-mHz tooth width is much narrower than the 122-kHz natural linewidth, 
the spectrum induced by a power of 40~$\mu$W recovers the natural lineshape of the corresponding ionic transition. 
The slightly broadened FWHM of 162~kHz is a consequence of power broadening which becomes more significant at the power of 4~mW with a 8.5-fold broadened FWHM of 1041~kHz. Nevertheless, measuring such spectra would enable the determination of the $^3P_1\rightarrow{}^1S_0$ transition energy in Ar$^{6+}$ to a relative accuracy of $\delta\omega/\omega=10^{-11}$, with an improvement by more than 6 orders of magnitude compared to current results~\cite{trigueiros1997atomic}.

For comb~2, whose 2.50-MHz tooth width is 20-fold broader than the natural linewidth, its excitation spectra depicted in Fig.~\ref{excitation_vs_detuning}b reveal the coherence properties of the comb itself. 
First, the spectrum induced by the 4-mW comb overestimates the tooth width by a factor of 2.2 due to power broadening, and predicts a relatively shorter coherence time of $29$~ns. 
However, with a 40-$\mu$W power, one obtains the lineshape of the comb tooth with a FWHM of $2.65\pm0.02$~MHz (the 0.02-MHz uncertainty is obtained from 100 realizations of the spectra), thus providing a good determination of the 64-ns coherence time with a 6\% deviation. 
Therefore, the temporal coherence of FCs can be verified on a time scale of several $\mu$s, which is orders of magnitude longer than in previous experiments~\cite{jones2005phase,yost2009vacuum,cingoz2012direct,corsi2017ultimate}. Furthermore, even for a comb coherence time as short as 64~ns, Fig.~\ref{excitation_vs_detuning}b shows that DFCS of the ions could still improve the accuracy of the transition energy by 5 orders of magnitude.

The verification of the 1-s-long coherence time of comb~1 would require tuning $\omega_0$ to the extremely narrow, 30-mHz, $^3P_2\rightarrow{}^1S_0$ forbidden transition around 14.331~eV. The effective Rabi frequencies of 35.6~Hz and 356~Hz for the EUV comb powers of 40 $\mu$W and 4 mW, respectively, would result in hundreds to thousands of Rabi cycles before the system evolves into a dynamical steady state, thus enabling full quantum control of the corresponding ionic states. The simulated FWHMs of the spectra, however, show widths of 132~Hz and 1.25~kHz for comb powers of 40~$\mu$W and 4~mW, respectively, due to power broadening. Though they are more than 3 orders of magnitude larger than the 30-mHz natural linewidth and the 160-mHz comb tooth width, they still represent an improvement in the accuracy of the $^3P_2\rightarrow{}^1S_0$ transition energy of Ar$^{6+}$ to the level of $\delta\omega/\omega=10^{-14}$, and set up the lower bound of the EUV comb coherence time to the range of milliseconds. Nevertheless, one can eliminate the power broadening to obtain a more accurate determination of the coherence time ((on the 10\% level in our current example, limited by the finite lifetime of the $^3P_2$ state) and of the transition energy by employing lower powers.

\textit{RFCS scheme --} 
Power broadening can be eliminated by implementing RFCS, where the ion is excited by two pulses separated from each other by $t_n=nT_{\rm{r}}$ ($n$ is the number of repetition cycles between the two pulses). When $t_n\ll\tau$, the total excitations by each pulse-pair can be calculated as~\cite{morgenweg2014ramsey} 
\begin{eqnarray}
\rho_{\text{ee}}(t_n)&=&\frac{|\mu|^2}{2}(\mathcal{E}_{\rm{eff}}T_{\rm{r}})^2\left\{1+\text{cos}[\Delta t_n+\phi(t_n)]\right\}.
\label{eq_rcsf}
\end{eqnarray} 
The first and second terms in the bracket of Eq.~(\ref{eq_rcsf}) describe the excitations resulting from the first and second pulse, respectively. Due to CEP dephasing, when $t_n$ becomes larger than $\tau_{\rm{c}}$, there is no deterministic and reproducible phase relation between the two pulses. Therefore, the averaged excitation reduces to~\cite{eramo2018analytic}
\begin{eqnarray}
\left<\rho_{\text{ee}}(t_n)\right>&=&\frac{|\mu|^2}{2}(\mathcal{E}_{\rm{eff}}T_{\rm{r}})^2\left[1+\text{cos}(\Delta t_n)e^{-\frac{1}{2}\sigma^2t_n}\right].
\label{average}
\end{eqnarray} 
While the cosine term generates Ramsey fringes and determines the ionic transition frequency~\cite{witte2005deep,morgenweg2014ramsey,altmann2018deep,dreissen2019high}, the exponentially decaying term determines the coherence time of the applied pulse train. Since the field strength appears in Eq.~(\ref{average}) as a prefactor, power broadening is eliminated in this case~\cite{morgenweg2014ramsey}. 
Therefore, RFCS of Ar$^{6+}$ ions can accurately measure the coherence time of the FC. Moreover, when the temporal coherence of comb~1 with $\tau_{\rm{c}}=1$~s is verified, it can also infer the corresponding transition frequency in Ar$^{6+}$ with $\delta\omega/\omega\sim10^{-17}$.

\textit{Conclusions --} 
We show that the implementation of the direct and Ramsey frequency comb spectroscopy of highly charged Mg-like ions can allow the determination of the coherence time of EUV FCs at time scales of several seconds, up to 7 orders of magnitude longer than in previous experiments, and improve the high-precision spectroscopy of EUV transitions by 12 orders of magnitude to the $\delta\omega/\omega\sim10^{-17}$ level. 
An experimental demonstration of these experiments will open the door to quantum control~\cite{adams2013x} of highly charged ions and enable applications in the search for physics beyond the standard model such as the variation of the fine-structure constant, the potential existence of a fifth force~\cite{safronova2018search}, and the electric dipole moment of elementary particles~\cite{chupp2019electric,kuchler2019searches}.


\begin{thebibliography}{59}
\expandafter\ifx\csname natexlab\endcsname\relax\def\natexlab#1{#1}\fi
\expandafter\ifx\csname bibnamefont\endcsname\relax
  \def\bibnamefont#1{#1}\fi
\expandafter\ifx\csname bibfnamefont\endcsname\relax
  \def\bibfnamefont#1{#1}\fi
\expandafter\ifx\csname citenamefont\endcsname\relax
  \def\citenamefont#1{#1}\fi
\expandafter\ifx\csname url\endcsname\relax
  \def\url#1{\texttt{#1}}\fi
\expandafter\ifx\csname urlprefix\endcsname\relax\def\urlprefix{URL }\fi
\providecommand{\bibinfo}[2]{#2}
\providecommand{\eprint}[2][]{\url{#2}}

\bibitem[{\citenamefont{Cundiff and Ye}(2003)}]{cundiff2003colloquium}
\bibinfo{author}{\bibfnamefont{S.~T.} \bibnamefont{Cundiff}} \bibnamefont{and}
  \bibinfo{author}{\bibfnamefont{J.}~\bibnamefont{Ye}}, \bibinfo{journal}{Rev.
  Mod. Phys.} \textbf{\bibinfo{volume}{75}}, \bibinfo{pages}{325}
  (\bibinfo{year}{2003}).

\bibitem[{\citenamefont{Ye and Cundiff}(2005)}]{ye2005femtosecond}
\bibinfo{author}{\bibfnamefont{J.}~\bibnamefont{Ye}} \bibnamefont{and}
  \bibinfo{author}{\bibfnamefont{S.~T.} \bibnamefont{Cundiff}},
  \emph{\bibinfo{title}{Femtosecond optical frequency comb: principle,
  operation and applications}} (\bibinfo{publisher}{Springer Science \&
  Business Media}, \bibinfo{year}{2005}).

\bibitem[{\citenamefont{Schibli et~al.}(2008)\citenamefont{Schibli, Hartl,
  Yost, Martin, Marcinkevi{\v{c}}ius, Fermann, and Ye}}]{schibli2008optical}
\bibinfo{author}{\bibfnamefont{T.}~\bibnamefont{Schibli}},
  \bibinfo{author}{\bibfnamefont{I.}~\bibnamefont{Hartl}},
  \bibinfo{author}{\bibfnamefont{D.}~\bibnamefont{Yost}},
  \bibinfo{author}{\bibfnamefont{M.}~\bibnamefont{Martin}},
  \bibinfo{author}{\bibfnamefont{A.}~\bibnamefont{Marcinkevi{\v{c}}ius}},
  \bibinfo{author}{\bibfnamefont{M.}~\bibnamefont{Fermann}}, \bibnamefont{and}
  \bibinfo{author}{\bibfnamefont{J.}~\bibnamefont{Ye}}, \bibinfo{journal}{Nat.
  Photonics} \textbf{\bibinfo{volume}{2}}, \bibinfo{pages}{355}
  (\bibinfo{year}{2008}).

\bibitem[{\citenamefont{Benko et~al.}(2012)\citenamefont{Benko, Ruehl, Martin,
  Eikema, Fermann, Hartl, and Ye}}]{benko2012full}
\bibinfo{author}{\bibfnamefont{C.}~\bibnamefont{Benko}},
  \bibinfo{author}{\bibfnamefont{A.}~\bibnamefont{Ruehl}},
  \bibinfo{author}{\bibfnamefont{M.}~\bibnamefont{Martin}},
  \bibinfo{author}{\bibfnamefont{K.}~\bibnamefont{Eikema}},
  \bibinfo{author}{\bibfnamefont{M.}~\bibnamefont{Fermann}},
  \bibinfo{author}{\bibfnamefont{I.}~\bibnamefont{Hartl}}, \bibnamefont{and}
  \bibinfo{author}{\bibfnamefont{J.}~\bibnamefont{Ye}}, \bibinfo{journal}{Opt.
  Lett.} \textbf{\bibinfo{volume}{37}}, \bibinfo{pages}{2196}
  (\bibinfo{year}{2012}).

\bibitem[{\citenamefont{Picqu{\'e} and H{\"a}nsch}(2019)}]{picque2019frequency}
\bibinfo{author}{\bibfnamefont{N.}~\bibnamefont{Picqu{\'e}}} \bibnamefont{and}
  \bibinfo{author}{\bibfnamefont{T.~W.} \bibnamefont{H{\"a}nsch}},
  \bibinfo{journal}{Nat. Photonics} \textbf{\bibinfo{volume}{13}},
  \bibinfo{pages}{146} (\bibinfo{year}{2019}).

\bibitem[{\citenamefont{Fortier and Baumann}(2019)}]{fortier201920}
\bibinfo{author}{\bibfnamefont{T.}~\bibnamefont{Fortier}} \bibnamefont{and}
  \bibinfo{author}{\bibfnamefont{E.}~\bibnamefont{Baumann}},
  \bibinfo{journal}{Communications Physics} \textbf{\bibinfo{volume}{2}},
  \bibinfo{pages}{1} (\bibinfo{year}{2019}).

\bibitem[{\citenamefont{Li et~al.}(2008)\citenamefont{Li, Benedick, Fendel,
  Glenday, K{\"a}rtner, Phillips, Sasselov, Szentgyorgyi, and
  Walsworth}}]{li2008laser}
\bibinfo{author}{\bibfnamefont{C.-H.} \bibnamefont{Li}},
  \bibinfo{author}{\bibfnamefont{A.~J.} \bibnamefont{Benedick}},
  \bibinfo{author}{\bibfnamefont{P.}~\bibnamefont{Fendel}},
  \bibinfo{author}{\bibfnamefont{A.~G.} \bibnamefont{Glenday}},
  \bibinfo{author}{\bibfnamefont{F.~X.} \bibnamefont{K{\"a}rtner}},
  \bibinfo{author}{\bibfnamefont{D.~F.} \bibnamefont{Phillips}},
  \bibinfo{author}{\bibfnamefont{D.}~\bibnamefont{Sasselov}},
  \bibinfo{author}{\bibfnamefont{A.}~\bibnamefont{Szentgyorgyi}},
  \bibnamefont{and} \bibinfo{author}{\bibfnamefont{R.~L.}
  \bibnamefont{Walsworth}}, \bibinfo{journal}{Nature}
  \textbf{\bibinfo{volume}{452}}, \bibinfo{pages}{610} (\bibinfo{year}{2008}).

\bibitem[{\citenamefont{Steinmetz et~al.}(2008)\citenamefont{Steinmetz, Wilken,
  Araujo-Hauck, Holzwarth, H{\"a}nsch, Pasquini, Manescau, D'Odorico, Murphy,
  Kentischer et~al.}}]{steinmetz2008laser}
\bibinfo{author}{\bibfnamefont{T.}~\bibnamefont{Steinmetz}},
  \bibinfo{author}{\bibfnamefont{T.}~\bibnamefont{Wilken}},
  \bibinfo{author}{\bibfnamefont{C.}~\bibnamefont{Araujo-Hauck}},
  \bibinfo{author}{\bibfnamefont{R.}~\bibnamefont{Holzwarth}},
  \bibinfo{author}{\bibfnamefont{T.~W.} \bibnamefont{H{\"a}nsch}},
  \bibinfo{author}{\bibfnamefont{L.}~\bibnamefont{Pasquini}},
  \bibinfo{author}{\bibfnamefont{A.}~\bibnamefont{Manescau}},
  \bibinfo{author}{\bibfnamefont{S.}~\bibnamefont{D'Odorico}},
  \bibinfo{author}{\bibfnamefont{M.~T.} \bibnamefont{Murphy}},
  \bibinfo{author}{\bibfnamefont{T.}~\bibnamefont{Kentischer}},
  \bibnamefont{et~al.}, \bibinfo{journal}{Science}
  \textbf{\bibinfo{volume}{321}}, \bibinfo{pages}{1335} (\bibinfo{year}{2008}).

\bibitem[{\citenamefont{Ludlow et~al.}(2015)\citenamefont{Ludlow, Boyd, Ye,
  Peik, and Schmidt}}]{ludlow2015optical}
\bibinfo{author}{\bibfnamefont{A.~D.} \bibnamefont{Ludlow}},
  \bibinfo{author}{\bibfnamefont{M.~M.} \bibnamefont{Boyd}},
  \bibinfo{author}{\bibfnamefont{J.}~\bibnamefont{Ye}},
  \bibinfo{author}{\bibfnamefont{E.}~\bibnamefont{Peik}}, \bibnamefont{and}
  \bibinfo{author}{\bibfnamefont{P.~O.} \bibnamefont{Schmidt}},
  \bibinfo{journal}{Rev. Mod. Phys.} \textbf{\bibinfo{volume}{87}},
  \bibinfo{pages}{637} (\bibinfo{year}{2015}).

\bibitem[{\citenamefont{Mills et~al.}(2012)\citenamefont{Mills, Hammond, Lam,
  and Jones}}]{mills2012xuv}
\bibinfo{author}{\bibfnamefont{A.~K.} \bibnamefont{Mills}},
  \bibinfo{author}{\bibfnamefont{T.}~\bibnamefont{Hammond}},
  \bibinfo{author}{\bibfnamefont{M.~H.} \bibnamefont{Lam}}, \bibnamefont{and}
  \bibinfo{author}{\bibfnamefont{D.~J.} \bibnamefont{Jones}},
  \bibinfo{journal}{J. Phys. B} \textbf{\bibinfo{volume}{45}},
  \bibinfo{pages}{142001} (\bibinfo{year}{2012}).

\bibitem[{\citenamefont{Gohle et~al.}(2005)\citenamefont{Gohle, Udem, Herrmann,
  Rauschenberger, Holzwarth, Schuessler, Krausz, and
  H{\"a}nsch}}]{gohle2005frequency}
\bibinfo{author}{\bibfnamefont{C.}~\bibnamefont{Gohle}},
  \bibinfo{author}{\bibfnamefont{T.}~\bibnamefont{Udem}},
  \bibinfo{author}{\bibfnamefont{M.}~\bibnamefont{Herrmann}},
  \bibinfo{author}{\bibfnamefont{J.}~\bibnamefont{Rauschenberger}},
  \bibinfo{author}{\bibfnamefont{R.}~\bibnamefont{Holzwarth}},
  \bibinfo{author}{\bibfnamefont{H.~A.} \bibnamefont{Schuessler}},
  \bibinfo{author}{\bibfnamefont{F.}~\bibnamefont{Krausz}}, \bibnamefont{and}
  \bibinfo{author}{\bibfnamefont{T.~W.} \bibnamefont{H{\"a}nsch}},
  \bibinfo{journal}{Nature} \textbf{\bibinfo{volume}{436}},
  \bibinfo{pages}{234} (\bibinfo{year}{2005}).

\bibitem[{\citenamefont{Jones et~al.}(2005)\citenamefont{Jones, Moll, Thorpe,
  and Ye}}]{jones2005phase}
\bibinfo{author}{\bibfnamefont{R.~J.} \bibnamefont{Jones}},
  \bibinfo{author}{\bibfnamefont{K.~D.} \bibnamefont{Moll}},
  \bibinfo{author}{\bibfnamefont{M.~J.} \bibnamefont{Thorpe}},
  \bibnamefont{and} \bibinfo{author}{\bibfnamefont{J.}~\bibnamefont{Ye}},
  \bibinfo{journal}{Phys. Rev. Lett.} \textbf{\bibinfo{volume}{94}},
  \bibinfo{pages}{193201} (\bibinfo{year}{2005}).

\bibitem[{\citenamefont{Cing{\"o}z et~al.}(2012)\citenamefont{Cing{\"o}z, Yost,
  Allison, Ruehl, Fermann, Hartl, and Ye}}]{cingoz2012direct}
\bibinfo{author}{\bibfnamefont{A.}~\bibnamefont{Cing{\"o}z}},
  \bibinfo{author}{\bibfnamefont{D.~C.} \bibnamefont{Yost}},
  \bibinfo{author}{\bibfnamefont{T.~K.} \bibnamefont{Allison}},
  \bibinfo{author}{\bibfnamefont{A.}~\bibnamefont{Ruehl}},
  \bibinfo{author}{\bibfnamefont{M.~E.} \bibnamefont{Fermann}},
  \bibinfo{author}{\bibfnamefont{I.}~\bibnamefont{Hartl}}, \bibnamefont{and}
  \bibinfo{author}{\bibfnamefont{J.}~\bibnamefont{Ye}},
  \bibinfo{journal}{Nature} \textbf{\bibinfo{volume}{482}}, \bibinfo{pages}{68}
  (\bibinfo{year}{2012}).

\bibitem[{\citenamefont{Dreissen et~al.}(2019)\citenamefont{Dreissen, Roth,
  Gr{\"u}ndeman, Krauth, Favier, and Eikema}}]{dreissen2019high}
\bibinfo{author}{\bibfnamefont{L.}~\bibnamefont{Dreissen}},
  \bibinfo{author}{\bibfnamefont{C.}~\bibnamefont{Roth}},
  \bibinfo{author}{\bibfnamefont{E.}~\bibnamefont{Gr{\"u}ndeman}},
  \bibinfo{author}{\bibfnamefont{J.}~\bibnamefont{Krauth}},
  \bibinfo{author}{\bibfnamefont{M.}~\bibnamefont{Favier}}, \bibnamefont{and}
  \bibinfo{author}{\bibfnamefont{K.}~\bibnamefont{Eikema}},
  \bibinfo{journal}{Phys. Rev. Lett.} \textbf{\bibinfo{volume}{123}},
  \bibinfo{pages}{143001} (\bibinfo{year}{2019}).

\bibitem[{\citenamefont{Adams et~al.}(2013)\citenamefont{Adams, Buth,
  Cavaletto, Evers, Harman, Keitel, P{\'a}lffy, Pic{\'o}n, R{\"o}hlsberger,
  Rostovtsev et~al.}}]{adams2013x}
\bibinfo{author}{\bibfnamefont{B.~W.} \bibnamefont{Adams}},
  \bibinfo{author}{\bibfnamefont{C.}~\bibnamefont{Buth}},
  \bibinfo{author}{\bibfnamefont{S.~M.} \bibnamefont{Cavaletto}},
  \bibinfo{author}{\bibfnamefont{J.}~\bibnamefont{Evers}},
  \bibinfo{author}{\bibfnamefont{Z.}~\bibnamefont{Harman}},
  \bibinfo{author}{\bibfnamefont{C.~H.} \bibnamefont{Keitel}},
  \bibinfo{author}{\bibfnamefont{A.}~\bibnamefont{P{\'a}lffy}},
  \bibinfo{author}{\bibfnamefont{A.}~\bibnamefont{Pic{\'o}n}},
  \bibinfo{author}{\bibfnamefont{R.}~\bibnamefont{R{\"o}hlsberger}},
  \bibinfo{author}{\bibfnamefont{Y.}~\bibnamefont{Rostovtsev}},
  \bibnamefont{et~al.}, \bibinfo{journal}{J. Mod. Opt.}
  \textbf{\bibinfo{volume}{60}}, \bibinfo{pages}{2} (\bibinfo{year}{2013}).

\bibitem[{\citenamefont{Cavaletto et~al.}(2014)\citenamefont{Cavaletto, Harman,
  Ott, Buth, Pfeifer, and Keitel}}]{cavaletto2014broadband}
\bibinfo{author}{\bibfnamefont{S.~M.} \bibnamefont{Cavaletto}},
  \bibinfo{author}{\bibfnamefont{Z.}~\bibnamefont{Harman}},
  \bibinfo{author}{\bibfnamefont{C.}~\bibnamefont{Ott}},
  \bibinfo{author}{\bibfnamefont{C.}~\bibnamefont{Buth}},
  \bibinfo{author}{\bibfnamefont{T.}~\bibnamefont{Pfeifer}}, \bibnamefont{and}
  \bibinfo{author}{\bibfnamefont{C.~H.} \bibnamefont{Keitel}},
  \bibinfo{journal}{Nature Photonics} \textbf{\bibinfo{volume}{8}},
  \bibinfo{pages}{520} (\bibinfo{year}{2014}).

\bibitem[{\citenamefont{Seiferle et~al.}(2019)\citenamefont{Seiferle, von~der
  Wense, Bilous, Amersdorffer, Lemell, Libisch, Stellmer, Schumm, D{\"u}llmann,
  P{\'a}lffy et~al.}}]{seiferle2019energy}
\bibinfo{author}{\bibfnamefont{B.}~\bibnamefont{Seiferle}},
  \bibinfo{author}{\bibfnamefont{L.}~\bibnamefont{von~der Wense}},
  \bibinfo{author}{\bibfnamefont{P.~V.} \bibnamefont{Bilous}},
  \bibinfo{author}{\bibfnamefont{I.}~\bibnamefont{Amersdorffer}},
  \bibinfo{author}{\bibfnamefont{C.}~\bibnamefont{Lemell}},
  \bibinfo{author}{\bibfnamefont{F.}~\bibnamefont{Libisch}},
  \bibinfo{author}{\bibfnamefont{S.}~\bibnamefont{Stellmer}},
  \bibinfo{author}{\bibfnamefont{T.}~\bibnamefont{Schumm}},
  \bibinfo{author}{\bibfnamefont{C.~E.} \bibnamefont{D{\"u}llmann}},
  \bibinfo{author}{\bibfnamefont{A.}~\bibnamefont{P{\'a}lffy}},
  \bibnamefont{et~al.}, \bibinfo{journal}{Nature}
  \textbf{\bibinfo{volume}{573}}, \bibinfo{pages}{243} (\bibinfo{year}{2019}).

\bibitem[{\citenamefont{Yost et~al.}(2009)\citenamefont{Yost, Schibli, Ye,
  Tate, Hostetter, Gaarde, and Schafer}}]{yost2009vacuum}
\bibinfo{author}{\bibfnamefont{D.~C.} \bibnamefont{Yost}},
  \bibinfo{author}{\bibfnamefont{T.~R.} \bibnamefont{Schibli}},
  \bibinfo{author}{\bibfnamefont{J.}~\bibnamefont{Ye}},
  \bibinfo{author}{\bibfnamefont{J.~L.} \bibnamefont{Tate}},
  \bibinfo{author}{\bibfnamefont{J.}~\bibnamefont{Hostetter}},
  \bibinfo{author}{\bibfnamefont{M.~B.} \bibnamefont{Gaarde}},
  \bibnamefont{and} \bibinfo{author}{\bibfnamefont{K.~J.}
  \bibnamefont{Schafer}}, \bibinfo{journal}{Nat. Phys.}
  \textbf{\bibinfo{volume}{5}}, \bibinfo{pages}{815} (\bibinfo{year}{2009}).

\bibitem[{\citenamefont{Benko et~al.}(2014)\citenamefont{Benko, Allison,
  Cing{\"o}z, Hua, Labaye, Yost, and Ye}}]{benko2014extreme}
\bibinfo{author}{\bibfnamefont{C.}~\bibnamefont{Benko}},
  \bibinfo{author}{\bibfnamefont{T.~K.} \bibnamefont{Allison}},
  \bibinfo{author}{\bibfnamefont{A.}~\bibnamefont{Cing{\"o}z}},
  \bibinfo{author}{\bibfnamefont{L.}~\bibnamefont{Hua}},
  \bibinfo{author}{\bibfnamefont{F.}~\bibnamefont{Labaye}},
  \bibinfo{author}{\bibfnamefont{D.~C.} \bibnamefont{Yost}}, \bibnamefont{and}
  \bibinfo{author}{\bibfnamefont{J.}~\bibnamefont{Ye}}, \bibinfo{journal}{Nat.
  Photonics} \textbf{\bibinfo{volume}{8}}, \bibinfo{pages}{530}
  (\bibinfo{year}{2014}).

\bibitem[{\citenamefont{Matei et~al.}(2017)\citenamefont{Matei, Legero,
  H{\"a}fner, Grebing, Weyrich, Zhang, Sonderhouse, Robinson, Ye, Riehle
  et~al.}}]{matei20171.5um}
\bibinfo{author}{\bibfnamefont{D.}~\bibnamefont{Matei}},
  \bibinfo{author}{\bibfnamefont{T.}~\bibnamefont{Legero}},
  \bibinfo{author}{\bibfnamefont{S.}~\bibnamefont{H{\"a}fner}},
  \bibinfo{author}{\bibfnamefont{C.}~\bibnamefont{Grebing}},
  \bibinfo{author}{\bibfnamefont{R.}~\bibnamefont{Weyrich}},
  \bibinfo{author}{\bibfnamefont{W.}~\bibnamefont{Zhang}},
  \bibinfo{author}{\bibfnamefont{L.}~\bibnamefont{Sonderhouse}},
  \bibinfo{author}{\bibfnamefont{J.}~\bibnamefont{Robinson}},
  \bibinfo{author}{\bibfnamefont{J.}~\bibnamefont{Ye}},
  \bibinfo{author}{\bibfnamefont{F.}~\bibnamefont{Riehle}},
  \bibnamefont{et~al.}, \bibinfo{journal}{Phys. Rev. Lett.}
  \textbf{\bibinfo{volume}{118}}, \bibinfo{pages}{263202}
  (\bibinfo{year}{2017}).

\bibitem[{\citenamefont{Corsi et~al.}(2017)\citenamefont{Corsi, Liontos,
  Bellini, Cavalieri, Pastor, de~Cumis, and Eramo}}]{corsi2017ultimate}
\bibinfo{author}{\bibfnamefont{C.}~\bibnamefont{Corsi}},
  \bibinfo{author}{\bibfnamefont{I.}~\bibnamefont{Liontos}},
  \bibinfo{author}{\bibfnamefont{M.}~\bibnamefont{Bellini}},
  \bibinfo{author}{\bibfnamefont{S.}~\bibnamefont{Cavalieri}},
  \bibinfo{author}{\bibfnamefont{P.~C.} \bibnamefont{Pastor}},
  \bibinfo{author}{\bibfnamefont{M.~S.} \bibnamefont{de~Cumis}},
  \bibnamefont{and} \bibinfo{author}{\bibfnamefont{R.}~\bibnamefont{Eramo}},
  \bibinfo{journal}{Phys. Rev. Lett.} \textbf{\bibinfo{volume}{118}},
  \bibinfo{pages}{143201} (\bibinfo{year}{2017}).

\bibitem[{\citenamefont{Eramo et~al.}(2018)\citenamefont{Eramo, Pastor, and
  Cavalieri}}]{eramo2018analytic}
\bibinfo{author}{\bibfnamefont{R.}~\bibnamefont{Eramo}},
  \bibinfo{author}{\bibfnamefont{P.~C.} \bibnamefont{Pastor}},
  \bibnamefont{and}
  \bibinfo{author}{\bibfnamefont{S.}~\bibnamefont{Cavalieri}},
  \bibinfo{journal}{Phys. Rev.~A} \textbf{\bibinfo{volume}{97}},
  \bibinfo{pages}{033842} (\bibinfo{year}{2018}).

\bibitem[{\citenamefont{Micke et~al.}(2018)\citenamefont{Micke, K{\"u}hn,
  Buchauer, Harries, B{\"u}cking, Blaum, Cieluch, Egl, Hollain, Kraemer
  et~al.}}]{micke2018heidelberg}
\bibinfo{author}{\bibfnamefont{P.}~\bibnamefont{Micke}},
  \bibinfo{author}{\bibfnamefont{S.}~\bibnamefont{K{\"u}hn}},
  \bibinfo{author}{\bibfnamefont{L.}~\bibnamefont{Buchauer}},
  \bibinfo{author}{\bibfnamefont{J.}~\bibnamefont{Harries}},
  \bibinfo{author}{\bibfnamefont{T.~M.} \bibnamefont{B{\"u}cking}},
  \bibinfo{author}{\bibfnamefont{K.}~\bibnamefont{Blaum}},
  \bibinfo{author}{\bibfnamefont{A.}~\bibnamefont{Cieluch}},
  \bibinfo{author}{\bibfnamefont{A.}~\bibnamefont{Egl}},
  \bibinfo{author}{\bibfnamefont{D.}~\bibnamefont{Hollain}},
  \bibinfo{author}{\bibfnamefont{S.}~\bibnamefont{Kraemer}},
  \bibnamefont{et~al.}, \bibinfo{journal}{Rev. Sci. Instrum.}
  \textbf{\bibinfo{volume}{89}}, \bibinfo{pages}{063109}
  (\bibinfo{year}{2018}).

\bibitem[{\citenamefont{Schwarz et~al.}(2012)\citenamefont{Schwarz, Versolato,
  Windberger, Brunner, Ballance, Eberle, Ullrich, Schmidt, Hansen, Gingell
  et~al.}}]{schwarz2012cryogenic}
\bibinfo{author}{\bibfnamefont{M.}~\bibnamefont{Schwarz}},
  \bibinfo{author}{\bibfnamefont{O.}~\bibnamefont{Versolato}},
  \bibinfo{author}{\bibfnamefont{A.}~\bibnamefont{Windberger}},
  \bibinfo{author}{\bibfnamefont{F.}~\bibnamefont{Brunner}},
  \bibinfo{author}{\bibfnamefont{T.}~\bibnamefont{Ballance}},
  \bibinfo{author}{\bibfnamefont{S.}~\bibnamefont{Eberle}},
  \bibinfo{author}{\bibfnamefont{J.}~\bibnamefont{Ullrich}},
  \bibinfo{author}{\bibfnamefont{P.~O.} \bibnamefont{Schmidt}},
  \bibinfo{author}{\bibfnamefont{A.~K.} \bibnamefont{Hansen}},
  \bibinfo{author}{\bibfnamefont{A.~D.} \bibnamefont{Gingell}},
  \bibnamefont{et~al.}, \bibinfo{journal}{Rev. Sci. Instrum.}
  \textbf{\bibinfo{volume}{83}}, \bibinfo{pages}{083115}
  (\bibinfo{year}{2012}).

\bibitem[{\citenamefont{Leopold et~al.}(2019)\citenamefont{Leopold, King,
  Micke, Bautista-Salvador, Heip, Ospelkaus, Crespo L{\'o}pez-Urrutia, and
  Schmidt}}]{leopold2019cryogenic}
\bibinfo{author}{\bibfnamefont{T.}~\bibnamefont{Leopold}},
  \bibinfo{author}{\bibfnamefont{S.~A.} \bibnamefont{King}},
  \bibinfo{author}{\bibfnamefont{P.}~\bibnamefont{Micke}},
  \bibinfo{author}{\bibfnamefont{A.}~\bibnamefont{Bautista-Salvador}},
  \bibinfo{author}{\bibfnamefont{J.~C.} \bibnamefont{Heip}},
  \bibinfo{author}{\bibfnamefont{C.}~\bibnamefont{Ospelkaus}},
  \bibinfo{author}{\bibfnamefont{J.~R.} \bibnamefont{Crespo
  L{\'o}pez-Urrutia}}, \bibnamefont{and} \bibinfo{author}{\bibfnamefont{P.~O.}
  \bibnamefont{Schmidt}}, \bibinfo{journal}{Rev. Sci. Instrum.}
  \textbf{\bibinfo{volume}{90}}, \bibinfo{pages}{073201}
  (\bibinfo{year}{2019}).

\bibitem[{\citenamefont{Micke et~al.}(2020)\citenamefont{Micke, Leopold, King,
  Benkler, Spie{\ss}, Schm{\"o}ger, Schwarz, Crespo L{\'o}pez-Urrutia, and
  Schmidt}}]{micke2020coherent}
\bibinfo{author}{\bibfnamefont{P.}~\bibnamefont{Micke}},
  \bibinfo{author}{\bibfnamefont{T.}~\bibnamefont{Leopold}},
  \bibinfo{author}{\bibfnamefont{S.~A.} \bibnamefont{King}},
  \bibinfo{author}{\bibfnamefont{E.}~\bibnamefont{Benkler}},
  \bibinfo{author}{\bibfnamefont{L.~J.} \bibnamefont{Spie{\ss}}},
  \bibinfo{author}{\bibfnamefont{L.}~\bibnamefont{Schm{\"o}ger}},
  \bibinfo{author}{\bibfnamefont{M.}~\bibnamefont{Schwarz}},
  \bibinfo{author}{\bibfnamefont{J.~R.} \bibnamefont{Crespo
  L{\'o}pez-Urrutia}}, \bibnamefont{and} \bibinfo{author}{\bibfnamefont{P.~O.}
  \bibnamefont{Schmidt}}, \bibinfo{journal}{Nature} p. \bibinfo{pages}{to be
  published} (\bibinfo{year}{2020}).

\bibitem[{\citenamefont{Nauta et~al.}(2017)\citenamefont{Nauta, Borodin, Ledwa,
  Stark, Schwarz, Schm{\"o}ger, Micke, Crespo L{\'o}pez-Urrutia, and
  Pfeifer}}]{nauta2017towards}
\bibinfo{author}{\bibfnamefont{J.}~\bibnamefont{Nauta}},
  \bibinfo{author}{\bibfnamefont{A.}~\bibnamefont{Borodin}},
  \bibinfo{author}{\bibfnamefont{H.~B.} \bibnamefont{Ledwa}},
  \bibinfo{author}{\bibfnamefont{J.}~\bibnamefont{Stark}},
  \bibinfo{author}{\bibfnamefont{M.}~\bibnamefont{Schwarz}},
  \bibinfo{author}{\bibfnamefont{L.}~\bibnamefont{Schm{\"o}ger}},
  \bibinfo{author}{\bibfnamefont{P.}~\bibnamefont{Micke}},
  \bibinfo{author}{\bibfnamefont{J.~R.} \bibnamefont{Crespo
  L{\'o}pez-Urrutia}}, \bibnamefont{and}
  \bibinfo{author}{\bibfnamefont{T.}~\bibnamefont{Pfeifer}},
  \bibinfo{journal}{Nucl. Instrum. Methods Phys. Res., Sect.~B}
  \textbf{\bibinfo{volume}{408}}, \bibinfo{pages}{285} (\bibinfo{year}{2017}).

\bibitem[{\citenamefont{Nauta et~al.}(2020)\citenamefont{Nauta, Oelmann,
  Ackermann, Knauer, Pappenberger, Borodin, Muhammad, Ledwa, Pfeifer, and
  Crespo L{\'o}pez-Urrutia}}]{nauta2020100MHz}
\bibinfo{author}{\bibfnamefont{J.}~\bibnamefont{Nauta}},
  \bibinfo{author}{\bibfnamefont{J.~H.} \bibnamefont{Oelmann}},
  \bibinfo{author}{\bibfnamefont{A.}~\bibnamefont{Ackermann}},
  \bibinfo{author}{\bibfnamefont{P.}~\bibnamefont{Knauer}},
  \bibinfo{author}{\bibfnamefont{R.}~\bibnamefont{Pappenberger}},
  \bibinfo{author}{\bibfnamefont{A.}~\bibnamefont{Borodin}},
  \bibinfo{author}{\bibfnamefont{I.~S.} \bibnamefont{Muhammad}},
  \bibinfo{author}{\bibfnamefont{H.}~\bibnamefont{Ledwa}},
  \bibinfo{author}{\bibfnamefont{T.}~\bibnamefont{Pfeifer}}, \bibnamefont{and}
  \bibinfo{author}{\bibfnamefont{J.~R.} \bibnamefont{Crespo
  L{\'o}pez-Urrutia}}, \bibinfo{journal}{Opt. Lett.}
  \textbf{\bibinfo{volume}{25}}, \bibinfo{pages}{75} (\bibinfo{year}{2020}).

\bibitem[{\citenamefont{Crespo L{\'o}pez-Urrutia}(2016)}]{lopez2016frequency}
\bibinfo{author}{\bibfnamefont{J.~R.} \bibnamefont{Crespo L{\'o}pez-Urrutia}},
  in \emph{\bibinfo{booktitle}{J. Phys. Conf. Ser.}}
  (\bibinfo{organization}{IOP Publishing}, \bibinfo{year}{2016}), vol.
  \bibinfo{volume}{723}, p. \bibinfo{pages}{012052}.

\bibitem[{\citenamefont{Kozlov et~al.}(2018)\citenamefont{Kozlov, Safronova,
  Crespo L{\'o}pez-Urrutia, and Schmidt}}]{kozlov2018highly}
\bibinfo{author}{\bibfnamefont{M.~G.} \bibnamefont{Kozlov}},
  \bibinfo{author}{\bibfnamefont{M.~S.} \bibnamefont{Safronova}},
  \bibinfo{author}{\bibfnamefont{J.~R.} \bibnamefont{Crespo
  L{\'o}pez-Urrutia}}, \bibnamefont{and} \bibinfo{author}{\bibfnamefont{P.~O.}
  \bibnamefont{Schmidt}}, \bibinfo{journal}{Rev. Mod. Phys.}
  \textbf{\bibinfo{volume}{90}}, \bibinfo{pages}{045005}
  (\bibinfo{year}{2018}).

\bibitem[{\citenamefont{Gustafsson et~al.}(2017)\citenamefont{Gustafsson,
  J{\"o}nsson, Fischer, and Grant}}]{gustafsson2017mcdhf}
\bibinfo{author}{\bibfnamefont{S.}~\bibnamefont{Gustafsson}},
  \bibinfo{author}{\bibfnamefont{P.}~\bibnamefont{J{\"o}nsson}},
  \bibinfo{author}{\bibfnamefont{C.~F.} \bibnamefont{Fischer}},
  \bibnamefont{and} \bibinfo{author}{\bibfnamefont{I.}~\bibnamefont{Grant}},
  \bibinfo{journal}{Astron. Astrophys.} \textbf{\bibinfo{volume}{597}},
  \bibinfo{pages}{A76} (\bibinfo{year}{2017}).

\bibitem[{\citenamefont{Fischer et~al.}(2019)\citenamefont{Fischer, Gaigalas,
  J{\"o}nsson, and Biero{\'n}}}]{fischer2019grasp2018}
\bibinfo{author}{\bibfnamefont{C.~F.} \bibnamefont{Fischer}},
  \bibinfo{author}{\bibfnamefont{G.}~\bibnamefont{Gaigalas}},
  \bibinfo{author}{\bibfnamefont{P.}~\bibnamefont{J{\"o}nsson}},
  \bibnamefont{and}
  \bibinfo{author}{\bibfnamefont{J.}~\bibnamefont{Biero{\'n}}},
  \bibinfo{journal}{Comput. Phys. Commun.} \textbf{\bibinfo{volume}{237}},
  \bibinfo{pages}{184} (\bibinfo{year}{2019}).

\bibitem[{\citenamefont{Kramida et~al.}(2019)\citenamefont{Kramida, Ralchenko,
  Reader, and {NIST ASD Team}}}]{kramida2019nist}
\bibinfo{author}{\bibfnamefont{A.}~\bibnamefont{Kramida}},
  \bibinfo{author}{\bibfnamefont{Y.}~\bibnamefont{Ralchenko}},
  \bibinfo{author}{\bibfnamefont{J.}~\bibnamefont{Reader}}, \bibnamefont{and}
  \bibinfo{author}{\bibnamefont{{NIST ASD Team}}}, \emph{\bibinfo{title}{{NIST}
  atomic spectra database (version 5.7. 1)}},
  \bibinfo{howpublished}{\url{https://physics.nist.gov/asd}}
  (\bibinfo{year}{2019}).

\bibitem[{\citenamefont{Cavalieri et~al.}(2002)\citenamefont{Cavalieri, Eramo,
  Materazzi, Corsi, and Bellini}}]{cavalieri2002ramsey}
\bibinfo{author}{\bibfnamefont{S.}~\bibnamefont{Cavalieri}},
  \bibinfo{author}{\bibfnamefont{R.}~\bibnamefont{Eramo}},
  \bibinfo{author}{\bibfnamefont{M.}~\bibnamefont{Materazzi}},
  \bibinfo{author}{\bibfnamefont{C.}~\bibnamefont{Corsi}}, \bibnamefont{and}
  \bibinfo{author}{\bibfnamefont{M.}~\bibnamefont{Bellini}},
  \bibinfo{journal}{Phys. Rev. Lett.} \textbf{\bibinfo{volume}{89}},
  \bibinfo{pages}{133002} (\bibinfo{year}{2002}).

\bibitem[{\citenamefont{Kandula et~al.}(2010)\citenamefont{Kandula, Gohle,
  Pinkert, Ubachs, and Eikema}}]{kandula2010extreme}
\bibinfo{author}{\bibfnamefont{D.~Z.} \bibnamefont{Kandula}},
  \bibinfo{author}{\bibfnamefont{C.}~\bibnamefont{Gohle}},
  \bibinfo{author}{\bibfnamefont{T.~J.} \bibnamefont{Pinkert}},
  \bibinfo{author}{\bibfnamefont{W.}~\bibnamefont{Ubachs}}, \bibnamefont{and}
  \bibinfo{author}{\bibfnamefont{K.~S.} \bibnamefont{Eikema}},
  \bibinfo{journal}{Phys. Rev. Lett.} \textbf{\bibinfo{volume}{105}},
  \bibinfo{pages}{063001} (\bibinfo{year}{2010}).

\bibitem[{\citenamefont{Witte et~al.}(2005)\citenamefont{Witte, Zinkstok,
  Ubachs, Hogervorst, and Eikema}}]{witte2005deep}
\bibinfo{author}{\bibfnamefont{S.}~\bibnamefont{Witte}},
  \bibinfo{author}{\bibfnamefont{R.~T.} \bibnamefont{Zinkstok}},
  \bibinfo{author}{\bibfnamefont{W.}~\bibnamefont{Ubachs}},
  \bibinfo{author}{\bibfnamefont{W.}~\bibnamefont{Hogervorst}},
  \bibnamefont{and} \bibinfo{author}{\bibfnamefont{K.~S.}
  \bibnamefont{Eikema}}, \bibinfo{journal}{Science}
  \textbf{\bibinfo{volume}{307}}, \bibinfo{pages}{400} (\bibinfo{year}{2005}).

\bibitem[{\citenamefont{Altmann et~al.}(2018)\citenamefont{Altmann, Dreissen,
  Salumbides, Ubachs, and Eikema}}]{altmann2018deep}
\bibinfo{author}{\bibfnamefont{R.}~\bibnamefont{Altmann}},
  \bibinfo{author}{\bibfnamefont{L.}~\bibnamefont{Dreissen}},
  \bibinfo{author}{\bibfnamefont{E.}~\bibnamefont{Salumbides}},
  \bibinfo{author}{\bibfnamefont{W.}~\bibnamefont{Ubachs}}, \bibnamefont{and}
  \bibinfo{author}{\bibfnamefont{K.}~\bibnamefont{Eikema}},
  \bibinfo{journal}{Phys. Rev. Lett.} \textbf{\bibinfo{volume}{120}},
  \bibinfo{pages}{043204} (\bibinfo{year}{2018}).

\bibitem[{\citenamefont{Felinto and L{\'o}pez}(2009)}]{felinto2009theory}
\bibinfo{author}{\bibfnamefont{D.}~\bibnamefont{Felinto}} \bibnamefont{and}
  \bibinfo{author}{\bibfnamefont{C.~E.} \bibnamefont{L{\'o}pez}},
  \bibinfo{journal}{Phys. Rev.~A} \textbf{\bibinfo{volume}{80}},
  \bibinfo{pages}{013419} (\bibinfo{year}{2009}).

\bibitem[{\citenamefont{Morgenweg et~al.}(2014)\citenamefont{Morgenweg, Barmes,
  and Eikema}}]{morgenweg2014ramsey}
\bibinfo{author}{\bibfnamefont{J.}~\bibnamefont{Morgenweg}},
  \bibinfo{author}{\bibfnamefont{I.}~\bibnamefont{Barmes}}, \bibnamefont{and}
  \bibinfo{author}{\bibfnamefont{K.~S.} \bibnamefont{Eikema}},
  \bibinfo{journal}{Nat. Phys.} \textbf{\bibinfo{volume}{10}},
  \bibinfo{pages}{30} (\bibinfo{year}{2014}).

\bibitem[{\citenamefont{Trigueiros et~al.}(1997)\citenamefont{Trigueiros,
  Mania, Gallardo, and Almandos}}]{trigueiros1997atomic}
\bibinfo{author}{\bibfnamefont{A.}~\bibnamefont{Trigueiros}},
  \bibinfo{author}{\bibfnamefont{A.}~\bibnamefont{Mania}},
  \bibinfo{author}{\bibfnamefont{M.}~\bibnamefont{Gallardo}}, \bibnamefont{and}
  \bibinfo{author}{\bibfnamefont{J.~R.} \bibnamefont{Almandos}},
  \bibinfo{journal}{J. Opt. Soc. Am. B} \textbf{\bibinfo{volume}{14}},
  \bibinfo{pages}{2463} (\bibinfo{year}{1997}).

\bibitem[{\citenamefont{Endo et~al.}(2018)\citenamefont{Endo, Shoji, and
  Schibli}}]{endo2018ultralow}
\bibinfo{author}{\bibfnamefont{M.}~\bibnamefont{Endo}},
  \bibinfo{author}{\bibfnamefont{T.~D.} \bibnamefont{Shoji}}, \bibnamefont{and}
  \bibinfo{author}{\bibfnamefont{T.~R.} \bibnamefont{Schibli}},
  \bibinfo{journal}{IEEE J. Sel. Top. Quantum Electron.}
  \textbf{\bibinfo{volume}{24}}, \bibinfo{pages}{1} (\bibinfo{year}{2018}).

\bibitem[{\citenamefont{Bartels et~al.}(2004)\citenamefont{Bartels, Oates,
  Hollberg, and Diddams}}]{bartels2004stabilization}
\bibinfo{author}{\bibfnamefont{A.}~\bibnamefont{Bartels}},
  \bibinfo{author}{\bibfnamefont{C.~W.} \bibnamefont{Oates}},
  \bibinfo{author}{\bibfnamefont{L.}~\bibnamefont{Hollberg}}, \bibnamefont{and}
  \bibinfo{author}{\bibfnamefont{S.~A.} \bibnamefont{Diddams}},
  \bibinfo{journal}{Opt. Lett.} \textbf{\bibinfo{volume}{29}},
  \bibinfo{pages}{1081} (\bibinfo{year}{2004}).

\bibitem[{\citenamefont{Liehl et~al.}(2019)\citenamefont{Liehl, Sulzer,
  Fehrenbacher, Rybka, Seletskiy, and Leitenstorfer}}]{liehl2019deterministic}
\bibinfo{author}{\bibfnamefont{A.}~\bibnamefont{Liehl}},
  \bibinfo{author}{\bibfnamefont{P.}~\bibnamefont{Sulzer}},
  \bibinfo{author}{\bibfnamefont{D.}~\bibnamefont{Fehrenbacher}},
  \bibinfo{author}{\bibfnamefont{T.}~\bibnamefont{Rybka}},
  \bibinfo{author}{\bibfnamefont{D.~V.} \bibnamefont{Seletskiy}},
  \bibnamefont{and}
  \bibinfo{author}{\bibfnamefont{A.}~\bibnamefont{Leitenstorfer}},
  \bibinfo{journal}{Phys. Rev. Lett.} \textbf{\bibinfo{volume}{122}},
  \bibinfo{pages}{203902} (\bibinfo{year}{2019}).

\bibitem[{\citenamefont{Temkin}(1993)}]{temkin1993excitation}
\bibinfo{author}{\bibfnamefont{R.}~\bibnamefont{Temkin}}, \bibinfo{journal}{J.
  Opt. Soc. Am.~B} \textbf{\bibinfo{volume}{10}}, \bibinfo{pages}{830}
  (\bibinfo{year}{1993}).

\bibitem[{\citenamefont{Ziolkowski et~al.}(1995)\citenamefont{Ziolkowski,
  Arnold, and Gogny}}]{ziolkowski1995ultrafast}
\bibinfo{author}{\bibfnamefont{R.~W.} \bibnamefont{Ziolkowski}},
  \bibinfo{author}{\bibfnamefont{J.~M.} \bibnamefont{Arnold}},
  \bibnamefont{and} \bibinfo{author}{\bibfnamefont{D.~M.} \bibnamefont{Gogny}},
  \bibinfo{journal}{Phys. Rev.~A} \textbf{\bibinfo{volume}{52}},
  \bibinfo{pages}{3082} (\bibinfo{year}{1995}).

\bibitem[{\citenamefont{Vitanov and Knight}(1995)}]{vitanov1995coherent}
\bibinfo{author}{\bibfnamefont{N.~V.} \bibnamefont{Vitanov}} \bibnamefont{and}
  \bibinfo{author}{\bibfnamefont{P.~L.} \bibnamefont{Knight}},
  \bibinfo{journal}{Phys. Rev.~A} \textbf{\bibinfo{volume}{52}},
  \bibinfo{pages}{2245} (\bibinfo{year}{1995}).

\bibitem[{\citenamefont{Scully and Zubairy}(1999)}]{scully1999quantum}
\bibinfo{author}{\bibfnamefont{M.~O.} \bibnamefont{Scully}} \bibnamefont{and}
  \bibinfo{author}{\bibfnamefont{M.~S.} \bibnamefont{Zubairy}},
  \emph{\bibinfo{title}{Quantum optics}} (\bibinfo{year}{1999}).

\bibitem[{\citenamefont{Porat et~al.}(2018)\citenamefont{Porat, Heyl, Schoun,
  Benko, D{\"o}rre, Corwin, and Ye}}]{porat2018phase}
\bibinfo{author}{\bibfnamefont{G.}~\bibnamefont{Porat}},
  \bibinfo{author}{\bibfnamefont{C.~M.} \bibnamefont{Heyl}},
  \bibinfo{author}{\bibfnamefont{S.~B.} \bibnamefont{Schoun}},
  \bibinfo{author}{\bibfnamefont{C.}~\bibnamefont{Benko}},
  \bibinfo{author}{\bibfnamefont{N.}~\bibnamefont{D{\"o}rre}},
  \bibinfo{author}{\bibfnamefont{K.~L.} \bibnamefont{Corwin}},
  \bibnamefont{and} \bibinfo{author}{\bibfnamefont{J.}~\bibnamefont{Ye}},
  \bibinfo{journal}{Nat. Photonics} \textbf{\bibinfo{volume}{12}},
  \bibinfo{pages}{387} (\bibinfo{year}{2018}).

\bibitem[{\citenamefont{Felinto et~al.}(2001)\citenamefont{Felinto, Bosco,
  Acioli, and Vianna}}]{felinto2001accumulative}
\bibinfo{author}{\bibfnamefont{D.}~\bibnamefont{Felinto}},
  \bibinfo{author}{\bibfnamefont{C.}~\bibnamefont{Bosco}},
  \bibinfo{author}{\bibfnamefont{L.}~\bibnamefont{Acioli}}, \bibnamefont{and}
  \bibinfo{author}{\bibfnamefont{S.}~\bibnamefont{Vianna}},
  \bibinfo{journal}{Phys. Rev.~A} \textbf{\bibinfo{volume}{64}},
  \bibinfo{pages}{063413} (\bibinfo{year}{2001}).

\bibitem[{\citenamefont{Felinto et~al.}(2003)\citenamefont{Felinto, Bosco,
  Acioli, and Vianna}}]{felinto2003coherent}
\bibinfo{author}{\bibfnamefont{D.}~\bibnamefont{Felinto}},
  \bibinfo{author}{\bibfnamefont{C.}~\bibnamefont{Bosco}},
  \bibinfo{author}{\bibfnamefont{L.}~\bibnamefont{Acioli}}, \bibnamefont{and}
  \bibinfo{author}{\bibfnamefont{S.}~\bibnamefont{Vianna}},
  \bibinfo{journal}{Opt. Commun.} \textbf{\bibinfo{volume}{215}},
  \bibinfo{pages}{69} (\bibinfo{year}{2003}).

\bibitem[{\citenamefont{Marian et~al.}(2004)\citenamefont{Marian, Stowe,
  Lawall, Felinto, and Ye}}]{marian2004united}
\bibinfo{author}{\bibfnamefont{A.}~\bibnamefont{Marian}},
  \bibinfo{author}{\bibfnamefont{M.~C.} \bibnamefont{Stowe}},
  \bibinfo{author}{\bibfnamefont{J.~R.} \bibnamefont{Lawall}},
  \bibinfo{author}{\bibfnamefont{D.}~\bibnamefont{Felinto}}, \bibnamefont{and}
  \bibinfo{author}{\bibfnamefont{J.}~\bibnamefont{Ye}},
  \bibinfo{journal}{Science} \textbf{\bibinfo{volume}{306}},
  \bibinfo{pages}{2063} (\bibinfo{year}{2004}).

\bibitem[{\citenamefont{Aumiler et~al.}(2009)\citenamefont{Aumiler, Ban, and
  Pichler}}]{aumiler2009time}
\bibinfo{author}{\bibfnamefont{D.}~\bibnamefont{Aumiler}},
  \bibinfo{author}{\bibfnamefont{T.}~\bibnamefont{Ban}}, \bibnamefont{and}
  \bibinfo{author}{\bibfnamefont{G.}~\bibnamefont{Pichler}},
  \bibinfo{journal}{Phys. Rev.~A} \textbf{\bibinfo{volume}{79}},
  \bibinfo{pages}{063403} (\bibinfo{year}{2009}).

\bibitem[{\citenamefont{Allen and Eberly}(1987)}]{allen1987optical}
\bibinfo{author}{\bibfnamefont{L.}~\bibnamefont{Allen}} \bibnamefont{and}
  \bibinfo{author}{\bibfnamefont{J.~H.} \bibnamefont{Eberly}},
  \emph{\bibinfo{title}{Optical resonance and two-level atoms}},
  vol.~\bibinfo{volume}{28} (\bibinfo{publisher}{Courier Corporation},
  \bibinfo{year}{1987}).

\bibitem[{\citenamefont{Moreno and Vianna}(2014)}]{moreno2014comparative}
\bibinfo{author}{\bibfnamefont{M.~P.} \bibnamefont{Moreno}} \bibnamefont{and}
  \bibinfo{author}{\bibfnamefont{S.~S.} \bibnamefont{Vianna}},
  \bibinfo{journal}{Opt. Commun.} \textbf{\bibinfo{volume}{313}},
  \bibinfo{pages}{113} (\bibinfo{year}{2014}).

\bibitem[{\citenamefont{Yudin et~al.}(2016)\citenamefont{Yudin, Taichenachev,
  and Basalaev}}]{yudin2016dynamic}
\bibinfo{author}{\bibfnamefont{V.}~\bibnamefont{Yudin}},
  \bibinfo{author}{\bibfnamefont{A.}~\bibnamefont{Taichenachev}},
  \bibnamefont{and} \bibinfo{author}{\bibfnamefont{M.~Y.}
  \bibnamefont{Basalaev}}, \bibinfo{journal}{Phys. Rev.~A}
  \textbf{\bibinfo{volume}{93}}, \bibinfo{pages}{013820}
  (\bibinfo{year}{2016}).

\bibitem[{\citenamefont{Cavaletto}(2014)}]{cavaletto2014Quantum}
\bibinfo{author}{\bibfnamefont{S.~M.} \bibnamefont{Cavaletto}},
  \emph{\bibinfo{title}{Quantum control of x-ray spectra}}
  (\bibinfo{publisher}{Ph.D thesis, Unviersity of Heidelberg, Germany},
  \bibinfo{year}{2014}).

\bibitem[{\citenamefont{Safronova et~al.}(2018)\citenamefont{Safronova, Budker,
  DeMille, Kimball, Derevianko, and Clark}}]{safronova2018search}
\bibinfo{author}{\bibfnamefont{M.}~\bibnamefont{Safronova}},
  \bibinfo{author}{\bibfnamefont{D.}~\bibnamefont{Budker}},
  \bibinfo{author}{\bibfnamefont{D.}~\bibnamefont{DeMille}},
  \bibinfo{author}{\bibfnamefont{D.~F.~J.} \bibnamefont{Kimball}},
  \bibinfo{author}{\bibfnamefont{A.}~\bibnamefont{Derevianko}},
  \bibnamefont{and} \bibinfo{author}{\bibfnamefont{C.~W.} \bibnamefont{Clark}},
  \bibinfo{journal}{Rev. Mod. Phys.} \textbf{\bibinfo{volume}{90}},
  \bibinfo{pages}{025008} (\bibinfo{year}{2018}).

\bibitem[{\citenamefont{Chupp et~al.}(2019)\citenamefont{Chupp, Fierlinger,
  Ramsey-Musolf, and Singh}}]{chupp2019electric}
\bibinfo{author}{\bibfnamefont{T.}~\bibnamefont{Chupp}},
  \bibinfo{author}{\bibfnamefont{P.}~\bibnamefont{Fierlinger}},
  \bibinfo{author}{\bibfnamefont{M.}~\bibnamefont{Ramsey-Musolf}},
  \bibnamefont{and} \bibinfo{author}{\bibfnamefont{J.}~\bibnamefont{Singh}},
  \bibinfo{journal}{Rev. Mod. Phys.} \textbf{\bibinfo{volume}{91}},
  \bibinfo{pages}{015001} (\bibinfo{year}{2019}).

\bibitem[{\citenamefont{Kuchler et~al.}(2019)}]{kuchler2019searches}
\bibinfo{author}{\bibfnamefont{F.}~\bibnamefont{Kuchler}} \bibnamefont{et~al.},
  \bibinfo{journal}{Universe} \textbf{\bibinfo{volume}{5}}, \bibinfo{pages}{56}
  (\bibinfo{year}{2019}).

\end{thebibliography}
\end{document}